\documentclass[12pt]{iopart}
\usepackage{bm}
\usepackage{graphicx}
\usepackage{amsfonts}
\begin{document}

\title{A simple topological model with continuous phase transition}

\author{F Baroni}

\ead{baronifab@libero.it}

\date{\today}

\begin{abstract}
In the area of topological and geometric treatment of phase transitions and symmetry breaking in Hamiltonian systems, in a recent paper some general sufficient conditions for these phenomena in $\mathbb{Z}_2$-symmetric systems (i.e. invariant under reflection of coordinates) have been found out. In this paper we present a simple topological model satisfying the above conditions hoping to enlighten the mechanism which causes this phenomenon in more general physical models. The symmetry breaking is testified by a continuous magnetization with a nonanalytic point in correspondence of a critical temperature which divides the broken symmetry phase from the unbroken one. A particularity with respect to the common pictures of a phase transition is that the nonanalyticity of the magnetization is not accompanied by a
nonanalytic behavior of the free energy.
\end{abstract}






\section{Introduction}

Phase transitions are sudden changes of the macroscopic behavior of a physical system composed by many interacting parts occurring while an external parameter is smoothly varied, generally the temperature, but e.g. in a quantum phase transition is the external magnetic field. The successful description of phase transitions starting from the properties of the microscopic interactions between the components of the system is one of the major achievements of equilibrium statistical mechanics.

From a statistical-mechanical point of view, in the canonical ensemble, a phase transition occurs at special values of the temperature $T$ called transition points, where thermodynamic quantities such as pressure, magnetization, or heat capacity, are non-analytic functions of $T$. These points are the boundaries between different phases of the system. Starting from the solution of the two-dimensional Ising model by Onsager \cite{onsager}, these singularities have been found in many models, and later developments like the renormalization group theory \cite{goldenfeld} have considerably deepened our knowledge of the properties of the transition points.

But in spite to the success of equilibrium statistical mechanics the issue of the deep origin of a phase transition remains open, and this motivates a study of phase transitions which may also be based on alternative approaches. One of such approaches, proposed in \cite{cccp} and developed later \cite{cpc}, is based on simple concepts and tools drawn from differential geometry and topology. The main issue of this new approach is a "topological hypothesis", whose content is that at their deepest level phase transitions in Hamiltonian systems are due to one or more topology changes of suitable submanifolds of configuration space, those where the system "lives" as the number of its degrees of freedom becomes very large.

This idea has been discussed and tested in many recent papers \cite{fcsp, aarz}. Moreover, the topological hypothesis has been given a rigorous background by a theorem \cite{fp} which states that, at least for systems with short-ranged interactions and confining potentials, topology changes in configuration space submanifolds are a necessary condition for a phase transition. However, the converse is not trivially true because there are models with topology changes without phase transitions \cite{ccp}.
The importance of the above theorem is to have established a strong link between phase transitions in Hamiltonian systems and the topology of suitable submanifolds of configuration space, and the main issue of the topological hypothesis is to search and to test some possible topology-based sufficient conditions for occurrence of phase transitions.

Indeed, in a recent paper \cite{bc} some sufficient topological conditions have been found out, although with the aid of some other conditions of geometric nature. Namely, for $\mathbb{Z}_2$-symmetric\footnote{$\mathbb{Z}_2$ is the group of integers modulo $2$ $\{0,1\}$ which are isomorphic to the symmetry group of the reflection of coordinates: $\textbf{q}\mapsto -\textbf{q}$.} systems  a theorem has been shown, according to which if the potential $V$ has two absolute minima separated by a minimum jump proportional to the number of degree of freedom $N$, then the system shows at least symmetry breaking \cite{goldenfeld,palmer}. Under suitable assumptions, the symmetry breaking can also be associated to a phase transition in the sense of loss of analyticity of the magnetization. In the same paper \cite{bc} also a very simple topological toy model, called "hypercubic model", has been built. This is a model with a first order symmetry breaking phase transition which shows in a pedagogical way how the theorem works.

This paper is devoted to the building of another topological toy model, that we call "hyperspherical model", showing $\mathbb{Z}_2$-symmetry breaking associated to continuous magnetization with a second order singularity in correspondence of a critical temperature. Despite this, the partition function does not show any singulary unlike what we expected. Indeed, this picture is quite anomalous because a singularity in the magnetization is generally associated to singularity also in the other thermodynamic functions, but this may be not so strange given the extreme simplicity of the model. It cannot expect to reproduce all the features of a physical model, e.g. Ising-like models. Nevertheless, the success in building up such toy models only by topological and geometrical ingredients may give some hints for enlightening how topology and geometry may affect the behavior of a general Hamiltonian system showing symmetry breaking and phase transitions.

Since we cannot presuppose an appropriate knowledge of the topological and geometric approach to symmetry breaking and phase transitions by the average reader, we devote \Sref{sec2} to a brief presentation of some basic concepts of it, and we refer to \cite{k} for a more exhaustive review. In \Sref{sec3} we recall some results about $\mathbb{Z}_2$-symmetry breaking obtained in \cite{bc}, comprised the aforementioned hypercubic model. Finally, in \Sref{sec4} we present the original results of this paper which is the aforementioned hyperspherical model showing $\mathbb{Z}_2$-symmetry breaking testified by a continuous magnetization.

\section{Basis of the \emph{topological hypothesis}}
\label{sec2}

All the particular properties of a Hamiltonian system have to be considered enclosed in the form of the potential $V$ in a cause-effect relationship, and topological hypothesis is an attempt to disclose this relationship for what concern phase transitions and symmetry breaking.

In particular, phase transitions show as loss of analyticity of the partition function $Z$ at some critical temperatures $T_c$, while symmetry breaking phenomena show by an order parameter which assumes a non vanishing value in a certain interval of values of temperatures $T$.

Generally, a phase transition and a symmetry breaking are associated and occur at a same critical temperature $T_c$. $T_c$ divides the broken symmetry phase from the unbroken one, although there are cases in which a system shows a phase transition without symmetry breaking. Topological hypothesis attempts to relate these phenomena to the topology of suitable submanifolds of configuration space.

Consider an $N$ degrees of freedom system with Hamiltonian given by
\begin{equation}
H(\textbf{p},\textbf{q})=\sum_{i=1}^N p_1^2+V(\textbf{q}).
\label{zc}
\end{equation}
The partition function $Z$ is
\begin{eqnarray}
Z(N,T)&=\int \rmd\mathbf{p}\,\rmd\mathbf{q}\,e^{-\frac{1}{T} H(\mathbf{p},\mathbf{q})}=\nonumber
\\
&=\int \rmd\mathbf{p}\,e^{-\frac{1}{T}\sum_{i=1}^N p_1^2}\int \rmd\mathbf{q}\,e^{-\frac{1}{T}V(\textbf{q})}=Z_{kin}Z_C
\end{eqnarray}
where $Z_{kin}$ is the kinetic part of $Z$, and $Z_C$ is the configurational part. In order to develop what follows we have to assume that the potential is lower bounded, and for convenience the minimum is assumed to be $0$. If $0$ is not the minimum it is sufficient to add to the potential a non-influential constant term equal to the minimum itself with the sign changed.

Consider $Z_C$, which can be decomposed as follows
\begin{eqnarray}
Z_C(N,T)=\int \rmd\mathbf{q}\,e^{-\frac{1}{T} V(\mathbf{q})}=N\int_{0}^{+\infty}\rmd v\,e^{-\frac{N v}{T}}\int_{\Sigma_v^N}\frac{\rmd\Sigma}{\vert\nabla V\vert}
\label{zc}
\end{eqnarray}
where $v=V/N$ is the potential per degree of freedom, and the $\Sigma_v^N$'s are the $v$-level sets of the potential $V$ in configuration space
\begin{equation}
\Sigma_v^N=\{\textbf{q}\in \mathbb{R}^N: V(\textbf{q})=N v\}.
\label{sigmav}
\end{equation}
The $\Sigma_v^N$'s are the boundary of the $M_v^N$'s ($\Sigma_v^N=\partial M_v^N$) defined as follows
\begin{equation}
M_v^N=\{\textbf{q}\in \mathbb{R}^N: V(\textbf{q})\leq N v\}.
\label{Mv}
\end{equation}

\eref{zc} shows that configuration space is foliated by the $\Sigma_v^N$'s by varying $v$ between $0$ and $+\infty$. The $\Sigma_v^N$'s are very important submanifolds because in the thermodynamic limit the canonical statistic measure narrows around $\Sigma_{\bar{v}(T)}^N$, where $\bar{v}(T)$ is the average potential per degree of freedom, and thus $\Sigma_{\bar{v}(T)}^N$ becomes the unique submanifold accessible to the representative point of the system.

This may have significant consequences on the one hand on the symmetries of the system and thus on the order parameter, and on the other hand on the analyticity of $Z_C$ in the thermodynamic limit, as the theorem in the following section suggests, owing to the fact that the $\Sigma_v^N$ have in general a very complex topology which changes by varying $v$.

The same considerations made about $Z_C$ can be made also for $Z_{kin}$, but in this case the corresponding submanifolds $\Sigma_t^N$, where $t=T/N$ is the kinetic energy per degree of freedom, are all trivially homeomorphic\footnote{A homeomorphism is a continuous bijection between manifolds with continuous inverse.} to the $N$-dim hypersphere. Further, $Z_{kin}$ is analytic for all values of $T$ in the thermodynamic limit, and thus cannot contribute to any loss of analyticity in $Z$.

\subsection{Necessary topological conditions for the occurrence of a phase transition}
\label{pettini_theorem}

The main result of the topological hypothesis so far obtained is a theorem which establishes a topological necessary condition for the occurrence of a phase transition in a Hamiltonian system with the potential of the standard form

\begin{equation}
V(\mathbf{q})=\sum_{i=1}^{N}\phi(q_i)+\sum_{i,j=1}^{N} c_{ij}\psi(|q_i-q_j|)
\end{equation}
which is short range, stable, confining and bounded below. We do not enter in the details of the theorem   because not essential for the following, so we refer the interested reader to \cite{fp}, \cite{fp1}. We limit to report the statement and a brief discussion of its main consequences:

\emph{Let be $v_{0}, v_{1}\in R$ such that $v_{0}<v_{1}$, if $\exists N_{0}:\forall N>N_{0},\forall v,v'\in [v_{0},v_{1}]$ $\Sigma_{v}^{N}\textrm{is diffeomorfic to}$ $\Sigma_{v'}^{N}$ then the limit for large $N$ of Helmholtz free energy $F$, is $C^{2}$ in the interval $(v_{0},v_{1})$, hence the system does not have any phase transition in the same interval al least of the second order.}
\\
\\
Having a phase transition in the interval $(v_{0},v_{1})$ means that the critical average potential $v^*=\bar{v}(T_c)$, where $T_c$ is the critical temperature, lies in it.

This theorem states a necessary condition for a phase transition, because if $v^*=\bar{v}(T_c)$ exists then the theorem implies that
\begin{equation}
\forall\,\epsilon>0,~\exists\,\overline{N}:~\forall\,N>\overline{N}\qquad\exists\,\Sigma_{v_{c}}^{N}:~|v_{c}-v^{*}|<\epsilon
\end{equation}
where $v_c$ is a value of the potential, generally different from $v^*$, at which a topological change occurs in the $\Sigma_{v_{c}}^{N}$. Thus it is possible to extract a sequence $\{\Sigma_{v_{c}^i}^{i}\}_{i\in N}$ such that $v_{c}^i\to v^*$ as $i\to\infty$, and in that limit we can say that the presence of a phase transition implies a topological change in the $\Sigma_{v}^{N}$'s located exactly in correspondence with $v^*$.

In the light of this theorem it is natural asking if its converse may hold, that is if a topological change in the $\Sigma_v^N$'s necessary causes a phase transition, but the answer is trivially not because it is very easy finding models with a lot of topological changes without phase transitions. Some of these models have been already studied, e.g. the $1$-dim $XY$ model \cite{ccp, mk}.

Further, in some other models the $\Sigma_v^N$'s undergoes huge topological changes that increases with $N$ and that not always are in correspondence with a phase transition, e.g. the mean-field $\phi^4$ model \cite{baroni} and the mean-field $XY$ model \cite{ccp1}. The difference between the last two models is that in the latter $v^*$ corresponds always to a topological change in the $\Sigma_{v}^{N}$'s for all values of the model's parameters, while in the former it is possible to find $v^*$ which does not correspond to any topological change in the $\Sigma_{v}^{N}$'s. There is no contradiction with the above theorem because among its hypothesis short range potential is requested, while in a mean-field model the interaction range is obviously infinite.

Further, it has been shown that in general no exclusively topological sufficient conditions for phase transitions are possible, we refer the reader to \cite{k1,k} for details.

\section{Toward sufficient topological and geometric conditions for symmetry breaking phase transitions}
\label{sec3}

In order to search for a sufficient topological and geometric condition for phase transitions it is necessary to study how topological changes in the $\Sigma_{v}^{N}$'s may affect the analytical properties of the thermodynamic functions. But in this work we will not follow this line of research, and we will shift our attention to the issue of how topology may break the symmetry of a system. In \cite{bc} a simple theorem on sufficient conditions for $\mathbb{Z}_2$-symmetry breaking, and an elementary models which illustrates how it works, have been found out.

Hereafter, we are going to present the trick necessary to perform the thermodynamic limit of $Z$, that is a necessary condition in order to occur symmetry breaking. We will see how this trick emphasizes the role of $\Sigma_{\bar{v}}^{N}$, and thus of its topology, in that limit.

The second integral on the right hand side of \eref{zc} is the $V$-derivative of the measure of the $M_v^N$'s induced by the standard metric of $R^N$ ($\rmd V=\rmd x_1\cdots \rmd x_N$)
\begin{equation}
\int_{\Sigma_v^N}\frac{\rmd\Sigma}{\vert\nabla V\vert}=\frac{\rmd}{N \rmd v}Mis(M_v^N)=\mu(\Sigma_v^N)
\end{equation}
that also coincides with the microcanonical volume of the $\Sigma_v^N$'s, $\mu(\Sigma_v^N)$. $Mis(M_v^N)$, being the volume of a subset of $\mathbb{R}^N$, can be rewritten as follows
\begin{equation}
Mis(M_v^N)=a_N(v)^N
\label{mismv}
\end{equation}
where the function $a_N(v)$ is defined. $a_N(v)$ has the dimension of a length, and is linked to the configurational entropy per degree of freedom $s_N(v)$ by the relation $s_N(v)=\ln a_N(v)$.

The introduction of $a_N(v)$ is useful in order to perform the thermodynamic limit of $Z_C$ by the saddle point trick. $\mu(\Sigma_v^N)$ can be written as a function of $a_N(v)$
\begin{equation}
\mu(\Sigma_v^N)=a'_N(v)a_N(v)^{N-1}
\end{equation}
where prime denotes the derivative with respect to $v$, and then we have
\begin{eqnarray}
Z_C=N\int_{0}^{+\infty}\rmd v\,e^{-\frac{Nv}{T}}\, a'_N(v)a_N(v)^{N-1}=N\int_{0}^{+\infty}\rmd v\frac{a'_N(v)}{a_N(v)}e^{-\frac{N}{T} f_N(v,T)}
\label{zcav}
\end{eqnarray}
where $f_N(v,T)=v-T\ln a_N(v)$ is the free energy per degree of freedom.

Now we can apply the saddle point trick to evaluate $Z_C$, but previously of performing the $N=\infty$ limit, it is necessary assuming that $\lim_{N\to\infty}a_N(v)=a(v)$ exists. This request corresponds to request that microcanonical entropy exists, which from a physical viewpoint seems quite reasonable. Thus
\begin{equation}
Z_C\simeq N\left(\frac{2\pi T}{N f''(\bar{v},T)}\right)^{\frac{1}{2}}\frac{a'(\bar{v})}{a(\bar{v})}e^{-\frac{N}{T}f(\bar{v},T)}
\end{equation}
where $f=\lim_{N\to\infty}f_N$, and $\bar{v}$ is the $v$-minimum of $f(v,T)$ in the interval $[0,+\infty)$ at fixed $T$. $\bar{v}$ coincides with the average potential per degree of freedom.

The saddle point trick implies that the configuration space accessible to the representative point $\textbf{q}$ of the system reduces to the $\Sigma_{\bar{v}}^N$ selected by $\bar{v}(T)$, and thus by $T$, in the limit of large $N$. Now we understand how the topological properties of the $\Sigma_{v}^{N}$'s may break the symmetry of a system.

\begin{figure}
\begin{center}
\includegraphics[width=8cm]{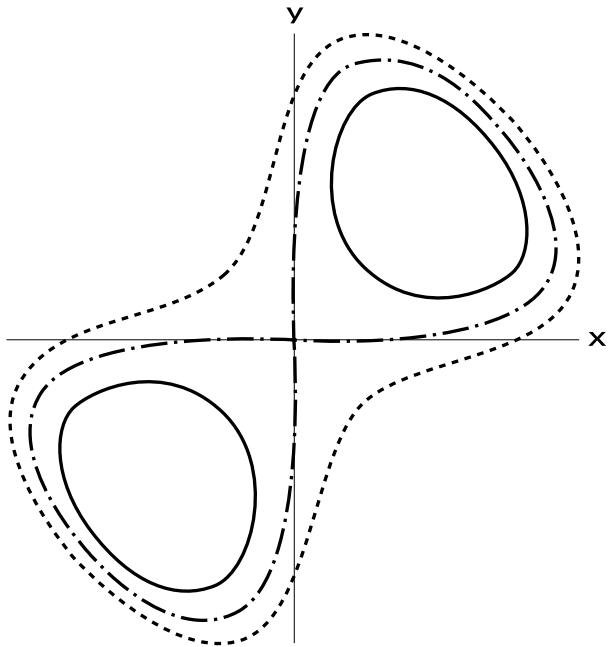}
\caption{Example of three $f$-level sets of the $\mathbb{Z}_2$-symmetric function $f=-\frac{1}{2}(q_1^2+q_2^2)+\frac{1}{4}(q_1^4+q_2^4)-q_1 q_2$, the $f$-levels are $-\frac{1}{2}, 0, 1$ respectively corresponding to the lines colors blue, magenta, yellow. The set labeled by the blue line has two connected components which do not conserve the $Z_2$-symmetry singularly.}
\label{fig_top}
\end{center}
\end{figure}

Indeed, suppose that the potential $V(\textbf{q})$ has some symmetries in configuration space, then the same symmetries have to belong also to the $\Sigma_{v}^{N}$'s which in general are composed by a number $n$ of connected components $\Sigma_{v}^{N, a}$ labeled by the index $a$, so the $\Sigma_{v}^{N}$ is the disjointed union of the $\Sigma_{v}^{N, a}$'s:
\begin{equation}
\Sigma_{v}^{N}=\bigcup_{a=1}^n\Sigma_{v}^{N, a}.
\end{equation}
The crucial observation is that a single $\Sigma_{v}^{N, a}$ does not need to have the same symmetries of the $\Sigma_{v}^{N}$ anymore, as the example in \Fref{fig_top} illustrates. In the light of this fact, in the thermodynamic limit the representative point $\textbf{q}$ can "live" only on a single c.c. $\Sigma_{\bar{v}(T)}^{N, a}$ because it cannot jump from one to the others anymore. In other words, the ergodicity cannot be assumed on the whole $\Sigma_{\bar{v}(T)}^N$ disregarding its topology.

This fact reflects on the magnetization $m_N$ per degree of freedom defined as
\begin{equation}
m_N=\frac{1}{Z_C}\int \rmd \textbf{q}\,\frac{1}{N}\sum_{i=1}^{N} q_i\,e^{-\frac{1}{T}V(\textbf{q})}
\label{mN}
\end{equation}
because in the limit of large $N$ the domain of integration has to be replaced by a single $\Sigma_{\bar{v}(T)}^{N, a}$, and thus the expression for $m_N$ has to be replaced with
\begin{equation}
m_N^a\simeq\frac{1}{\mu\left(\Sigma_{\bar{v}(T)}^{N, a}\right)}\int_{\Sigma_{\bar{v}(T)}^{N, a}}\rmd\Sigma\frac{1}{N}\sum_{i=1}^{N} q_i\frac{1}{\vert\nabla V\vert}.
\label{m}
\end{equation}
In (\ref{m}) we have assumed the ergodicity on the $\Sigma_{\bar{v}(T)}^{N, a}$. $\{m_N^a\}_{a=1,\dots,n}$ are the set of all possible values of the magnetization which undergoes to a splitting every time that $\Sigma_{\bar{v}(T)}^{N}$ undergoes to a topological change by varying $T$. The temperatures at which the topological changes occur are recognized as critical temperatures.

At this point it must be noted that in the thermodynamic limit other selection mechanisms may intervene to limit the ergodicity of a system on the $\Sigma_{\bar{v}(T)}^{N}$ besides the topological one here pointed out. For instance, consider the mean-field $\phi^4$ model \cite{baroni}. For suitable values of the parameters and of the temperature it shows $\mathbb{Z}_2$-symmetry breaking with  $\Sigma_{\bar{v}(T)}^{N}$ homeomorphic to the $N$-dim hypersphere, and thus in this case the topology of the $\Sigma_{\bar{v}(T)}^{N}$ cannot be the acting selection mechanism.

\subsection{Sufficient topological and geometric conditions for $\mathbb{Z}_2$-symmetry breaking}
\label{baroni_theorem}

Hereafter we will deal only with systems having $\mathbb{Z}_2$-symmetry, i.e. with potential $V(\textbf{q})$ invariant under reflection of coordinates $\textbf{q}\rightarrow -\textbf{q}$. Considerations of the previous section can be applied to these system and condensed in a straightforward theorem on necessary conditions for $Z_2$-symmetry breaking \cite{bc}.

Statement:

\emph{Let us consider a Hamiltonian system with $N$ degree of freedom and the potential $V$ bounded below which is $Z_2$-symmetric. Let the entropy per degree of freedom be well defined in the
thermodynamic limit, i.e., the function $a(v)$ defined in equation (\ref{mismv}) exist and be continuous
and piecewise differentiable. Let $\Sigma_v^N$ be the family of equipotential submanifolds of the
configuration space defined as in equation (\ref{sigmav}). Without loss of generality, let $min(V )=0$.
\\
Let $v''>v'>0$ be two values of the potential energy per degree of freedom $V/N$ such
that $\forall\, v: \,v\ge v'', \forall N$ $\Sigma_v^N$ is made of a single connected component, and such that $\forall \, v: v'>v\ge 0, \forall N$ $\Sigma_v^N$ is made of more than one connected component which are not $Z_2$-symmetric singularly considered.
\\
Then, in the thermodynamic limit the $Z_2$-symmetry is spontaneously broken for all the temperatures $T<T'$ where $T'$ is such that $v'=\overline{v}(T')$, and is unbroken for all the temperatures $T\ge T''$ where $T''$ is such that $v''=\overline{v}(T'')$, provided the $\Sigma_v^N$'s remain ergodic in the thermodynamic limit.}
\\
\\
This theorem also implies the occurrence of a singularity in the order parameter because it has to vanish for $T\ge T''$, and has to be not vanishing for $T\le T'$, but since it is not possible joining in an analytical way the null function with a not null one, necessarily a loss of analyticity must occurs for at least a critical temperature $T_c$ such that $T'\le T_c\le T''$. $T_c$ corresponds to a critical value of the average potential $v_c=\bar{v}(T_c)$ such that $v'\le v_c\le v''$. Further, if we restrict the condition of the theorem in a such way that $v''=v'$ then the singularity is located exactly in correspondence of the critical potential $v_c=v''=v'$, and if $T''=T'$ then $T_c=T''=T'$.

It must be noted that it might be very hard to show if the potential $V(\mathbf{q})$ of a general physical model satisfies the assumptions of the theorem, because finding out the topology of the $\Sigma_v^N$'s is generally a very difficult task. Despite this, it cannot be excluded that future developments of the research in this field could provide suitable tools. For now, we are content to see at work the theorem in two elementary models, one of which is briefly recalled in the next subsection, and the other is the original part of this paper.

\subsection{The \emph{hypercubic model} with $\mathbb{Z}_2$-symmetry breaking and first order phase transition}

\begin{figure}
\begin{center}
\includegraphics[width=6cm]{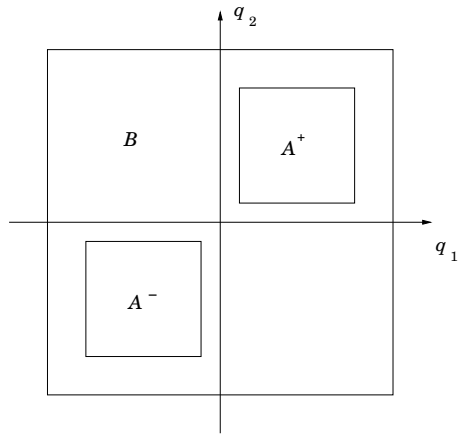}
\caption{Sketch of the hypercubes $A^{\pm}$ and $B$ for $N = 2$. From \cite{bc}.}
\label{hc}
\end{center}
\end{figure}

Now we describe briefly a topological model given in \cite{bc} to enlighten, in a pedagogical way, the content of the theorem of the last subsection. We build a double-hole potential $V$ which is $\mathbb{Z}_2$-symmetric by using $N$-dimensional hypercubes
\begin{equation}
V(\textbf{q})=\left\{\begin{array}{ll}
0\quad\quad\,\, &\hbox{if} \quad\textbf{q}\in A^{\pm}
\\
N v_c &\hbox{if} \quad\textbf{q}\in B\backslash\{A^+\cup A^- \}
\\
+\infty &\hbox{if} \quad\textbf{q}\in \mathbb{R}^N\backslash B.
\end{array}\right.
\end{equation}
The configuration space is $\mathbb{R}^N$, $A^+$ and $A^-$ are two disjoint hypercubes not centered in the origin and symmetric under $\mathbb{Z}_2$, and $B$ is a hypercube centered in the origin such that $A^+\cup A^- \subset B$. \Fref{hc} can help to understand the disposal. Note that by construction the minimum jump to pass from one hole to the other is proportional to $N$, this assumption is essential to make this model satisfying the hypothesis of the theorem.

The $\Sigma_v^N$'s are the following

\begin{equation}
\Sigma_v^N=\left\{\begin{array}{ll}
\emptyset \qquad\qquad\qquad\quad &\hbox{if} \quad v<0
\\
A^+\cup A^- &\hbox{if} \quad v=0
\\
\emptyset &\hbox{if} \quad 0<v<v_c
\\
B\backslash\{A^+\cup A^- \} &\hbox{if} \quad v=v_c
\\
\emptyset &\hbox{if} \quad v>v_c
\end{array}\right.
\end{equation}
from which we see that the permitted values of the potential are only two: $0$ and $v_c$. The partition function $Z_N$ is

\begin{equation}
Z_N(T)=2 a^N+(b^N-2a^N)\,e^{-\frac{N v_c}{T}}.
\end{equation}
where $a$ and $b$ are the sides of the hypercubes $A^{\pm}$ and $B$ respectively.

We called this model \emph{hypercubic model} (the name \emph{hypercubic model} is only conventional, because the hypercubes can be substituted by other geometric figures), which satisfies the hypothesis of the theorem in the last section with $v''=v_c$, $v'=0$. Indeed, in the limit $T\to 0$ $\Sigma_0^N$ is selected, and in the limit $T\to \infty$ $\Sigma_{v_c}^N$ is selected. Then we expect the occurrence of $\mathbb{Z}_2$-symmetry breaking associated to a first order singularity in the magnetization for at least a finite critical temperature $T_c$.

\begin{figure}
\begin{center}
\includegraphics[angle=270, width=8cm]{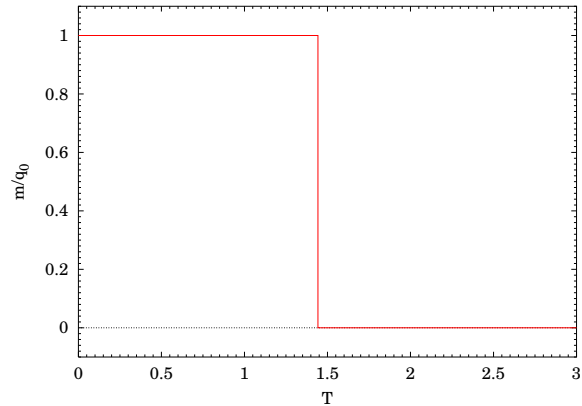}
\caption{Magnetization per degree of freedom vs T as $N\to\infty$ of the hypercubic model as $a=1$, $b=2a$, $v_c=1$, and so $T_c=1/\ln 2$. From \cite{bc}.}
\label{hcmT}
\end{center}
\end{figure}

Indeed, the analytical solution in the thermodynamic limit shows that this is just the case with $T_c=v_c \ln ^{-1}(b/a)$. The magnetization per degree of freedom in the broken symmetry phase is simply the coordinate of the center of mass of $A^+$ or $A^-$, the picture is sketched in \Fref{hcmT}. Even though not implicated by the theorem, a singularity at $T_c$ occurs also in the partition function and thus in the thermodynamic functions, reproducing so the common picture of a first order phase transition.

\begin{figure}
\begin{center}
\includegraphics[angle=270, width=8cm]{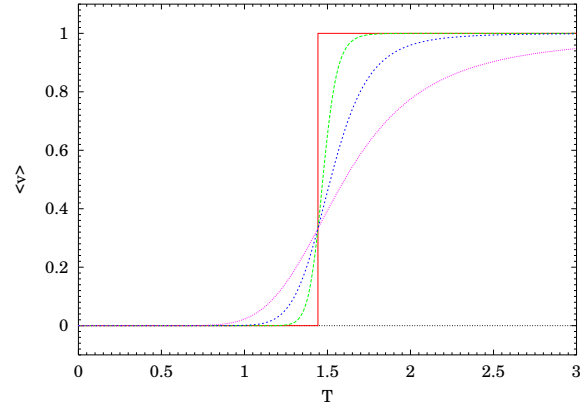}
\caption{Average potential of the hypercubic model vs T. The different
smooth curves are the finite-$N$ result with $N = 10, 20, 50$, while the piecewise constant
curve is the $N\to\infty$ limit. Numerical values as in \Fref{hcmT}. From \cite{bc}.}
\label{hcvT}
\end{center}
\end{figure}

It is worth remarking that in building of the hypercubic model, instead of hypercubes, we can use any other manifolds, provided they are topologically equivalent, e.g. hyperspheres.

Now we make some observations on the relation between symmetry breaking and the singularity in the average potential. The structure of the hypercubic model implies that the former needs the latter, but the converse is not true. Indeed, we can redefine $\Sigma_0^N$ by only one hypercube $A$ centered in the origin of coordinates and with the same side $a$, so that the $\mathbb{Z}_2$-symmetry never break, but the solution of the thermodynamic does not change in the thermodynamic limit, and so the critical temperature $T_c$ occurs the same.

\section{The \emph{hyperspherical model}}
\label{sec4}

In this section we build a topological model, called \emph{hyperspherical model}, showing $\mathbb{Z}_2$-symmetry breaking with a continuous magnetization which passes by a nonanalytic point. This point separates the broken symmetry phase to the unbroken one. The basic ingredients are $N$-dim hyperballs\footnote{With hyperball we mean a hypersphere with its interior.} by which we will build the $M_v^N$'s of the potential as defined in \eref{Mv}.

Generally, when a potential is defined the starting point is its explicit expression $V(\textbf{q})$, but here we define directly the $M_v^N$'s because we are interested in how their topology and measure match to entail a phase transition, and not in the explicit expression of $V$ itself. This is simply possible under the condition that if $v'<v''$ then $M_{v'}\subseteq M_{v''}$. Nevertheless, it is not excluded that the potential we are going to define may be also given by an explicit expression, but it may be very complicated and it would be useless for our present purposes.

Let us fix a $v_c>0$. For $v\ge v_c$ let $M_v^N$ be an $N$-dim hyperball $B_v$ centered in the origin of coordinates such that $Mis(B_v)=v^N$, while for $v_c>v\ge 0$ let $M_v^N$ be the disjointed union of two $N$-dim semi-hyperballs $B_v^+$, $B_v^-$. $B_v^+$, $B_v^-$ are obtained by dividing one $N$-dim hyperball centered in the origin by the hyperplane $\sum_{i=i}^{N}q_i=0$, and by widening $B_v^+$ and $B_v^-$ thus obtained. So constructed, $B_v^+$ and $B_v^-$ are the images of each other under $Z_2$-symmetry. Further, we assume $Mis(B_v^+\cup B_v^-)=v^N$.

Resuming, at the varying of $v$ from $+\infty$ to $0$ $M_v^N$ is an $N$-dim hyperball with measure $v^N$ which disconnects in two connected components for a critical value $v_c$

\begin{equation}
M_v^N=\left\{\begin{array}{ll}
B_v\qquad\qquad &\hbox{if} \quad v\ge v_c
\\
B_v^+\cup B_v^-  &\hbox{if} \quad v_c>v\ge 0.
\end{array}\right.
\end{equation}

The last step in building the model is to choose how to widen $B_v^+$ and $B_v^-$ by translating them parallel to their axes of symmetry. We can choose between two extreme cases: to do not widen $B_v^+$ and $B_v^-$ at all, letting them in touch, or to widen $B_v^+$ and $B_v^-$ maximally in such a way that all the $B_v^+$'s have the pole in common for $v_c\ge v\ge 0$. The same for the $B_v^-$'s.

These choices reflect on the shape of the magnetization, which will be vanishing without symmetry breaking at all for the first choice, or which will reach its maximum value for the second choice. In the following we will consider the second choice.

\begin{figure}
\begin{center}
\includegraphics[width=8cm]{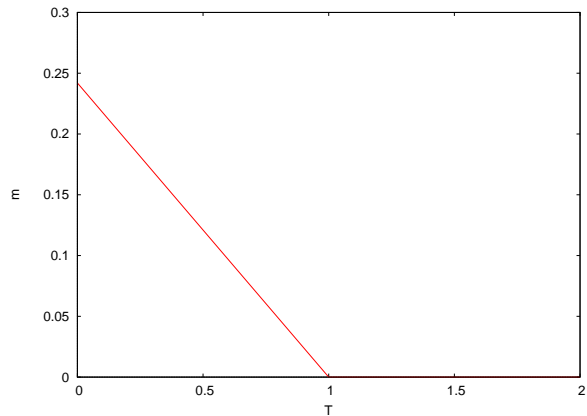}
\caption{Magnetization per degree of freedom vs $T$ for the hyperspherical model (red line). $v_c=1$ is assumed, and thus $T_c=v_c=1$. The dashed line labels another pattern of the magnetization among the other infinite possible ones in the broken symmetry phase corresponding to a difference choice of arranging the $M_v$'s as $v<v_c$, as explained in the text.}
\label{fig_mT}
\end{center}
\end{figure}

Obviously, infinite intermediate possible disposals exist, dictated only by the request that the potential $V$ is a single value function of the coordinate $\textbf{q}$'s. In these all intermediate cases the only constraint on the shape of the magnetization is such that its slope is bounded below by the tangent of the maximum magnetization sketched in \Fref{fig_mT} and upper bounded by $0$.

Now we have a look at what the $\Sigma_v^N$'s \eref{sigmav} are, we recall that they are the boundary of the $M_v^N$'s:  $\Sigma_v^N=\partial M_v^N$. For $v>v_c$ $\Sigma_v^N$ is the boundary of $B_v$, called $S_v$, that is an $N$-dim hypersphere of the same radius. For $v_c>v\ge 0$ $\Sigma_v^N$ is the boundary of $B_v^+\cup B_v^-$, called $S_v^+\cup S_v^-$, that is two $N$-dim semi-hyperspheres each closed by one $(N-1)$-dim hypersphere of the same radius. Finally, for $v=v_c$ $\Sigma_{v_c}^N$, called $S_{v_c}$, is an $N$-dim hypersphere jointed with an $(N-1)$-dim hypersphere of the same radius. Summarizing

\begin{equation}
\Sigma_v^N=\left\{\begin{array}{ll}
S_v \quad\quad\quad\,\,\, &\hbox{if} \quad v>v_c
\\
S_{v_c} &\hbox{if} \quad v=v_c
\\
S_v^+\cup S_v^-  &\hbox{if} \quad v_c>v\ge 0.
\end{array}\right.
\end{equation}
$S_{v_c}$ plays the role of the critical $v$-level by which the disconnection from one connected component of the $\Sigma_v^N$'s to two occurs.

Now we resolve the thermodynamic. The function $a_N(v)$, given by definition \eref{mismv}, is independent on $N$ and is

\begin{equation}
a(v)=v\quad \hbox{if}\quad v\ge 0.
\label{av}
\end{equation}
The configurational partition function $Z_C$ is given by \eref{zcav}
\begin{equation}
Z_C=N\int_{0}^{+\infty}\rmd v\, e^{-\frac{Nv}{T}}\, v^{N-1}=\frac{N!}{N^{N}}T^N
\end{equation}
disregarding terms depending on $N$ only, the free energy per degree of freedom is

\begin{equation}
f(T)=-\frac{T}{N}\ln Z_C=-T\ln T
\label{fT}
\end{equation}
the average potential per degree of freedom and the specific heat are respectively

\begin{equation}
\overline{v}(T)=-T^2\frac{\partial}{\partial T}\left(\frac{f}{T}\right)=T
\label{vT}
\end{equation}
\begin{eqnarray}
c_v(T)=\frac{\partial\overline{v}}{\partial T}=1.
\end{eqnarray}
Note that they do not depend on $N$, and therefore it is not necessary performing the thermodynamic limit. At $T=v_c$ the disconnection of the $\Sigma_v^N$ occurs, and thus we define $T_c=v_c$ the critical temperature.

\begin{figure}
\begin{center}
\includegraphics[width=8cm]{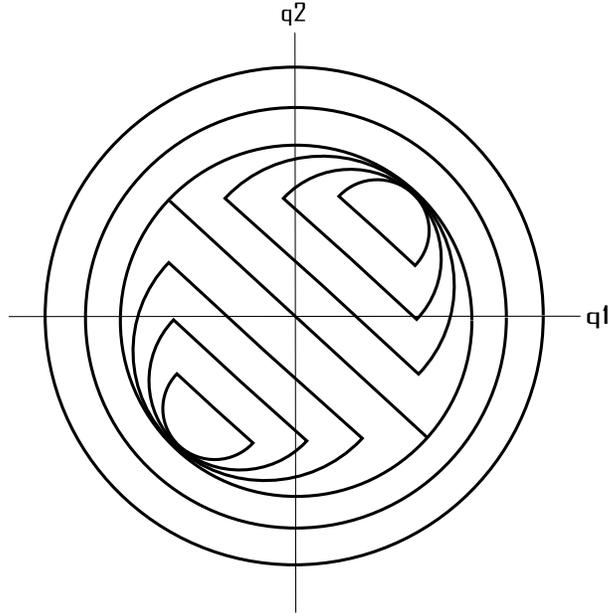}
\caption{Some $\Sigma_v^N$'s at $N=2$ in the configuration space $(q_1,q_2)$ for the hyperspherical model.}
\label{fig6_sigmav}
\end{center}
\end{figure}
\begin{figure}
\begin{center}
\includegraphics[width=8cm]{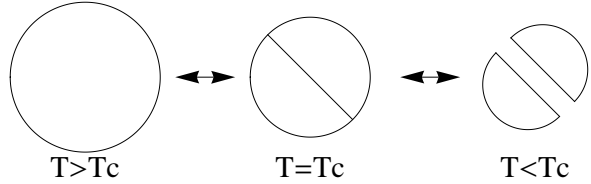}
\caption{Some $\Sigma_{\bar{v}(T)}^N$'s of the hyperspherical models schematically sketched at $N=2$ which illustrate the topological mechanism of $\mathbb{Z}_2$-symmetry breaking.}
\label{fig_mectop}
\end{center}
\end{figure}

Now we pass to study the magnetization per degree of freedom $m_N$ defined in \eref{m}. We start finding out the average of the representative point $\langle\textbf{q}\rangle$ on the $\Sigma_v^N$'s at finite $N$, and then performing the limit $N\to\infty$. The radius of the hyperspheres that constitutes the $M_v^N$'s is fixed by the formula of the volume of the $N$-dim hypersphere
\begin{equation}
a(v)^N=\frac{\pi^{\frac{N}{2}}}{\frac{N}{2}\Gamma\left(\frac{N}{2}\right)} R_N(v)^N
\label{rmv}
\end{equation}
and by the definition \eref{av} of $a(v)$. As $v\ge v_c$, $\langle\mathbf{q}\rangle$ vanishes on the $S_v$'s for reasons of symmetry. As $v_c>v\ge 0$ we have to calculate $\langle\mathbf{q}\rangle$ on the $S_v^+$'s or on the $S_v^-$'s, thus we need the knowledge of $(\vert\nabla V\vert)^{-1}$ on them, but owing of their properties of symmetry we can perform the same the calculation without the detailed knowledge of $(\vert\nabla V\vert)^{-1}$.

Indeed, because of our choice of arranging the $S_v^+$'s and the $S_v^-$'s, $(\vert\nabla V\vert)^{-1}$ vanishes at the pole, and it is constant on the parallels by an increasing value as the parallel goes from the pole to the equator, where it reaches its maximum. This maximum is the same on the $(N-1)$-dim hyperball which closes the $N$-dim semi-hypersphere. From this we can deduce that $\langle\textbf{q}\rangle$ is a point lying on the axis of symmetry of the $S_v^+$'s or $S_v^-$'s, and lying between the center of mass of the $N$-dim semi-hypersphere and the center of the corresponding whole $N$-dim hypersphere.

Now we focus our attention to the center of mass of the $N$-dim semi-hypersphere. We fix a coordinate system with the origin coinciding with the center of the corresponding hypersphere. The axis $x_N$ coincides with the axis of symmetry of the $N$-dim semi-hypersphere, and the other axes are orthonormal to $x_N$. Obviously, for reasons of symmetry only the component along $x_N$ is not vanishing, and in Appendix \ref{aA} we show that its value is
\begin{equation}
B_N=\frac{2 R}{\sqrt\pi N}
\end{equation}
where $R$ is the radius.

Returning to $\langle\textbf{q}\rangle=\left(\langle q_1\rangle,\dots,\langle q_N\rangle\right)$, it is linked to the magnetization per degree of freedom $m_N$ by the relation
\begin{equation}
m_N=\frac{1}{N}\sum_{i=1}^N\langle q_i\rangle
\end{equation}
but since $\forall i,j$ $\langle q_i\rangle=\langle q_j\rangle$ for reasons of symmetry, it follows that
\begin{equation}
|\langle\textbf{q}\rangle|^2=\sum_{i=1}^N\langle q_i\rangle^2=N m_N^2.
\label{qm}
\end{equation}
As $0\leq v<v_c$
\begin{equation}
R_N(v_c)-R_N(v)< |\langle\textbf{q}\rangle| <R_N(v_c)-R_N(v)+B_N(v)
\label{mq}
\end{equation}
where $R_N(v)$ is the radius of $B_v^{N+}$ or $B_v^{N-}$. Then, by using \eref{rmv}, \eref{qm} and \eref{mq} we obtain
\begin{eqnarray}
\frac{1}{\sqrt{N}}\left(\frac{ N\Gamma\left(\frac{N}{2}\right)}{2\pi^{\frac{N}{2}}}\right)^{\frac{1}{N}}\left(a(v_c)-a(v)\right)<m_N(v)<\nonumber
\\
<\frac{1}{\sqrt{N}}\left(\frac{ N\Gamma\left(\frac{N}{2}\right)}{2\pi^{\frac{N}{2}}}\right)^{\frac{1}{N}}\left(a(v_c)-a(v)\left(1-\frac{2}{\sqrt{\pi}{N}}\right)\right)
\end{eqnarray}
by performing the limit $N\to\infty$, and by using the relations $\lim_{N\to\infty}N^{\frac{1}{N}}=1$, $\lim_{N\to\infty}\frac{N!^{\frac{1}{N}}}{N}=\frac{1}{e}$ we have
\begin{equation}
m(v)=\lim_{N\to\infty}m_N(v)=\frac{1}{\sqrt{2\pi e}}(a(v_c)-a(v)).
\end{equation}
Finally, by using \eref{av} and \eref{vT}, we reconstruct the link with the temperature $T$
\begin{equation}
m(T)=\left\{\begin{array}{ll}
0\quad\qquad\qquad\quad &\hbox{if}\quad T\ge T_c
\\
\frac{1}{\sqrt{2\pi e}}(v_c-T) &\hbox{if}\quad T\leq T_c
\end{array}\right.
\end{equation}
where, we recall, $T_c=v_c$.

This model satisfies the assumptions of the theorem in \Sref{baroni_theorem} with $v''=v'=v_c$, $T''=T'=T_c$, thus we had to expect exactly what the analytical solution shows: $Z_2$-symmetry breaking associated to a singularity in the magnetization exactly located at $T=T_c$. But this does not reproduce the usual picture of a symmetry breaking phase transition because it is not accompanied with any singularity in the partition function $Z$, and thus in the thermodynamic functions. This is quite surprising, and until now we have not been able to understand why this is the case.

\section{Concluding remarks and outlooks}

In the book \cite{pettini} the author, among a lot of other things, points out the strong relation between phase transitions and symmetry breaking and the topology of the $v$-level sets of the potential $\Sigma_v^N$'s \eref{sigmav} in $N$ degrees of freedom Hamiltonian systems. However, the question if topology might be involved in the deep origin of those phenomena remains substantially open, although some recent results \cite{k1} show the impossibility of purely topological sufficient conditions.

In \cite{bc} an attempt to deepen this issue has been made showing a straightforward theorem, reported in \Sref{baroni_theorem}, on topological and geometric sufficient conditions for $\mathbb{Z}_2$-symmetry breaking, which points out the importance of the topology of the $\Sigma_{\bar{v}(T)}^N$ in the thermodynamic limit selected by the temperature $T$.

Indeed, in that limit the canonical measure narrows more and more around $\Sigma_{\bar{v}(T)}^N$, and since the latter is generally made by more than one connected components which do not need to be $\mathbb{Z}_2$-symmetric, then the representative point has to choose among them and thus $\mathbb{Z}_2$-symmetry can be broken.

The original part of this work has consisted in the construction of a topological model, called "hyperspherical model", which illustrates how the above mentioned theorem works. The $\Sigma_v^N$'s are directly defined in terms of hyperspheres (hence the name of the model) which disconnects in two connected components below a critical temperature $T_c$.

The magnetization is continuous and shows a second order singularity in correspondence of $T_c$. Despite this, the partition function shows no singularity, and thus the thermodynamic functions too. That is quite surprising for a model with symmetry breaking, but its extreme abstractness does not guarantee a realistic reproduction of the properties of a physical model, e.g. Ising-like models.

Anyway, the aim of this toy model, as well as the hypercubic one, is highlighting how the topological mechanism of selection among distinct connected components of the $\Sigma_v^N$'s works, although it is not the unique possible selection mechanism able to induce symmetry breaking. For example, in the mean-field $\phi^4$ model \cite{baroni} we have found out that, for some values of the parameters and the temperature, the $\Sigma_v^N$'s are homeomorphic to a hypersphere also in the broken symmetry phase, and thus another selection mechanism has to act.

However, this fact does not exclude that topological mechanism might be at the origin of symmetry breaking any case, because it might be applied to other subsets of configuration space more constrained with respect to the $\Sigma_v^N$'s. Indeed, there is no reason to assume the ergodicity on the $\Sigma_{\bar{v}(T)}^N$ in the limit of large $N$, we have simply assumed it both in the theorem and in the hyperspherical model.

\ack

I would like to thank Lapo Casetti for his fundamental help in developing the results in this work.

\appendix

\section{The center of mass of the $N$-dim semi-hypersphere}
\label{aA}

Let a coordinate system be with the origin coinciding with the center of the $N$-dim hypersphere of radius $R$. The axis $x_N$ is coinciding with its axis of symmetry, and the other axes are orthogonal to $x_N$. Because of the symmetry of the semi-hypersphere, only the component of the center of mass along $x_N$, $B_N$, is not vanishing, and thus we limit to calculate only it.

We start with the $N=3$ case, and then we will generalize the result by induction to a general $N$
\begin{equation}
B_3=\frac{2 R}{4\pi}\int_{0}^{\pi}\rmd\theta_2\sin^2\theta_2\int_{0}^{\pi}\rmd\theta_1\sin\theta_1
\end{equation}
where $(\theta_1,\theta_2)$ are standard polar coordinates, and $R$ is the radius. The generalization is straightforward
\begin{eqnarray}
B_N=\frac{2 R}{Mis(\mathbb{S}^{N-1})}\int_{0}^{\pi}\rmd\theta_{N-1}\sin^{N-1}\theta_{N-1}\cdots\int_{0}^{\pi}\rmd\theta_{1}\sin\theta_{1}\nonumber
\end{eqnarray}
where $\mathbb{S}^N$ denoted the $N$-dim hypersphere of unitary radius. By using the following formula

\begin{eqnarray}
Mis(\mathbb{S}^{N-1})=\frac{2\pi^{\frac{N}{2}}} {\Gamma\left(\frac{N}{2}\right)}
\\
\int_{0}^{\pi}\rmd\theta_{1}\sin^N\theta=\frac{\sqrt{\pi}\Gamma\left(\frac{N}{2}+\frac{1}{2}\right)}   {\Gamma\left(\frac{1}{2}\right)}
\\
\Gamma(1)=0!=1
\\
\Gamma(x+1)=x\Gamma(x)
\end{eqnarray}
and by some trivial algebraic manipulation the searched result is
\begin{equation}
B_N=\frac{2 R}{\sqrt\pi N}.
\end{equation}

\end{document}